\input{aipcheck}

\documentclass[
    ,final            
    ,sort                
  ]
  {aipproc}

\layoutstyle{6x9}

\begin{document}

\title{The EXOTIME targets HS\,0702+6043 and HS\,0444+0458}



\author{Ronny Lutz}{
  address={Georg-August Universit\"at G\"ottingen, 
    Institut f\"ur Astrophysik,\\ Friedrich-Hund-Platz 1, 37077 G\"ottingen, Germany}
  ,altaddress={IMPRS on Solar System Physics, Max-Planck Institute for Solar System Research, 
    Max-Planck-Stra\ss e 2, 37191 Katlenburg-Lindau, Germany}
}

\author{Sonja Schuh}{
  address={Georg-August Universit\"at G\"ottingen, 
    Institut f\"ur Astrophysik,\\ Friedrich-Hund-Platz 1, 37077 G\"ottingen, Germany}
}

\author{Roberto Silvotti}{
  address={Osservatorio Astronomico di Torino, strada dell'Osservatorio 20, 10025 Pino Torinese, Italy}
}

\begin{abstract}
Pulsations in subdwarf B (sdB) stars are an important tool to constrain the
evolutionary status of these evolved objects. Interestingly, the same data used
for this asteroseismological approach can also be used to search for substellar
companions around these objects by analyzing the timing of the pulsations
by means of a so-called O--C diagram. Substellar objects around sdB stars are 
important for two different reasons: they are suspected to be able to influence the 
evolution of their host-star and they are an ideal test case to examine the properties of 
exoplanets which have survived the red giant expansion of their host stars.
\end{abstract}

\maketitle


\section{Introduction}
\begin{figure}
  \includegraphics[height=.4\textheight]{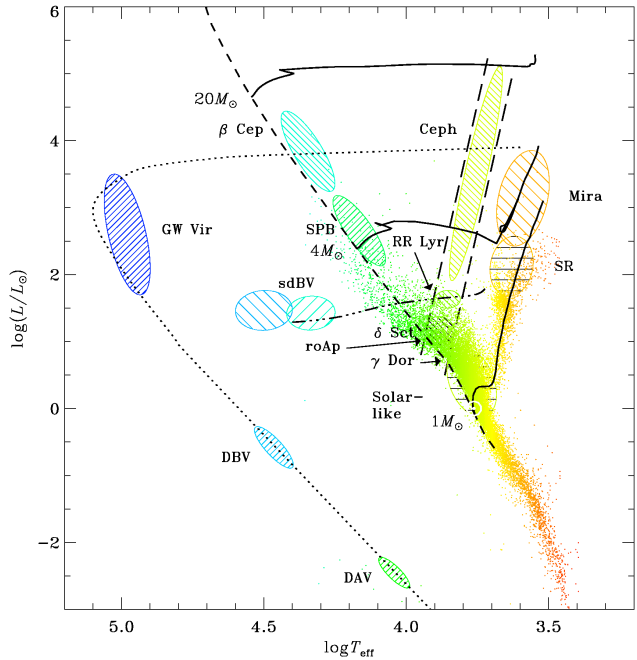}
  \caption{Pulsational Hertzsprung-Russell diagram showing the location
  of the variable subdwarf B stars (sdBV). Modified version from 
  http://astro.phys.au.dk/HELAS/puls\_HR.
  }
  \label{fig:1}
\end{figure}

Hot subdwarf B (sdB) stars are evolved objects with canonical masses close
to half a Solar mass. They are burning helium in their core and populate
the so-called extreme horizontal branch (EHB). During their 
evolution on the Red Giant branch, the progenitors must suffer an extreme
mass loss, such that at the time of the helium flash there is only a very 
thin hydrogen envelope left. The thin envelope is inert and prevents the
sdB from shell burning and therefore from an evolution towards the AGB
and PN phases. Instead, these objects are believed to directly evolve to
the White Dwarf cooling sequence (see Fig. \ref{fig:1}).
There is also the possibility that sdB stars show pulsations driven by
a $\kappa$~mechanism related to an opacity bump caused by iron group elements
\citep{1997ApJ...483L.123C}. Two different types of modes can appear: short period
pressure \mbox{(p-)} modes or long period gravity (g-) modes. For both groups,
the rapidly pulsating sdBV$_r$ and the slowly pulsating sdBV$_s$ stars, the
amplitudes can be as low as only some mmag. There also exist a few hybrid
objects (sdBV$_{rs}$) belonging to both groups
\citep{2006A&A...445L..31S, 2009A&A...496..469L}.
\par
Examining the pulsations of sdB stars provides the observational input
for asteroseismological modeling, in order to characterize the internal structure.
Furthermore, a measured change in the period, $\dot{P}$, can additionally
constraint asteroseismological models and give 
hints on mode identification. Measuring $\dot{P}$, and therefore determining the evolutionary 
timescale, is one of our goals.
\par
The search for exoplanets around sdB stars is another goal of our project:
more than 490 exoplanets are known to date. Due to the applicability of the
common detection methods as well as the specific goal to find an Earth-analogon in a 
solar-system-like exo-system, most of the exoplanet surveys are focused on 
solar-like stars on the Main Sequence. However, little is known about the late stages 
of exoplanetary system evolution, in particular the fate of exoplanets during and after 
the Red Giant expansion of the host star. The number of exoplanets around Giant and Subgiant 
stars is increasing, but still low compared to the Main Sequence hosts. The same is true
for White Dwarfs and for hot subdwarfs. The latter are the type of star that we aim
to consider here. Finding exoplanets around sdB stars might be immediate proof
that exoplanets would be able to survive the host star expansion or even a common
envelope phase.
\par
The formation of sdB stars is not very well understood: the binary sdB population
can quite well be explained by the current binary formation channels including
common envelope evolution and stable and unstable RLOF \citep{2002MNRAS.336..449H},
however there are issues to explain apparently single sdB stars. One of
the more promising scenarios for single sdB formation is given by
\citet{1998AJ....116.1308S}, who examines the possibility 
of enhanced mass loss of the sdB progenitor during the Red Giant phase induced
by substellar companions. Therefore, finding substellar
objects around single sdB stars would directly support this
hypothesis.
\par
We aim to apply a timing method (O--C analysis, observed minus calculated) to search for 
substellar companions around pulsating sdB stars. At the same time, with the same data, we 
are able to derive the evolutionary timescales for these objects. Different shapes of an
O--C diagram can be interpreted as follows: a parabolic trend indicates a linear change
in the pulsation period which is under consideration. From this quadratic O--C, one can
calculate a value $\dot{P}$, i.e. the change of the pulsation period with time. This
$\dot{P}$ can then be translated to an evolutionary timescale $T_{evol}$ of the pulsating 
sdB star via
\begin{equation}
T_{evol} = P/\dot{P}.
\end{equation}
In addition, finding sinusoidal residuals can most plausibly be explained by an orbital
reflex motion of the pulsating star, where the timing signals used to
construct the \mbox{O--C} diagram in our case are the times of maximum intensity.
Concerning sdB stars, the object HS\,2201+2610
is the only object where an O--C analysis has been published yet. \citet{2007Natur.449..189S} find an 
evolutionary timescale of 7.6 Myr and also the signal of an exoplanetary companion with a 
minimum mass of 3.2 times Jupiter's mass in an orbit of 1170 days. This exoplanet is the first 
one having been detected around a sdB star. If this object formed as a first generation planet at 
the same time as the host star, it would be direct proof that the substellar object survived the 
Red Giant expansion of its host star. Triggered by this discovery, we
set up the \emph{EXOTIME} program (see next section).
\begin{ltxfigure}
  \includegraphics[width=.499\textwidth]{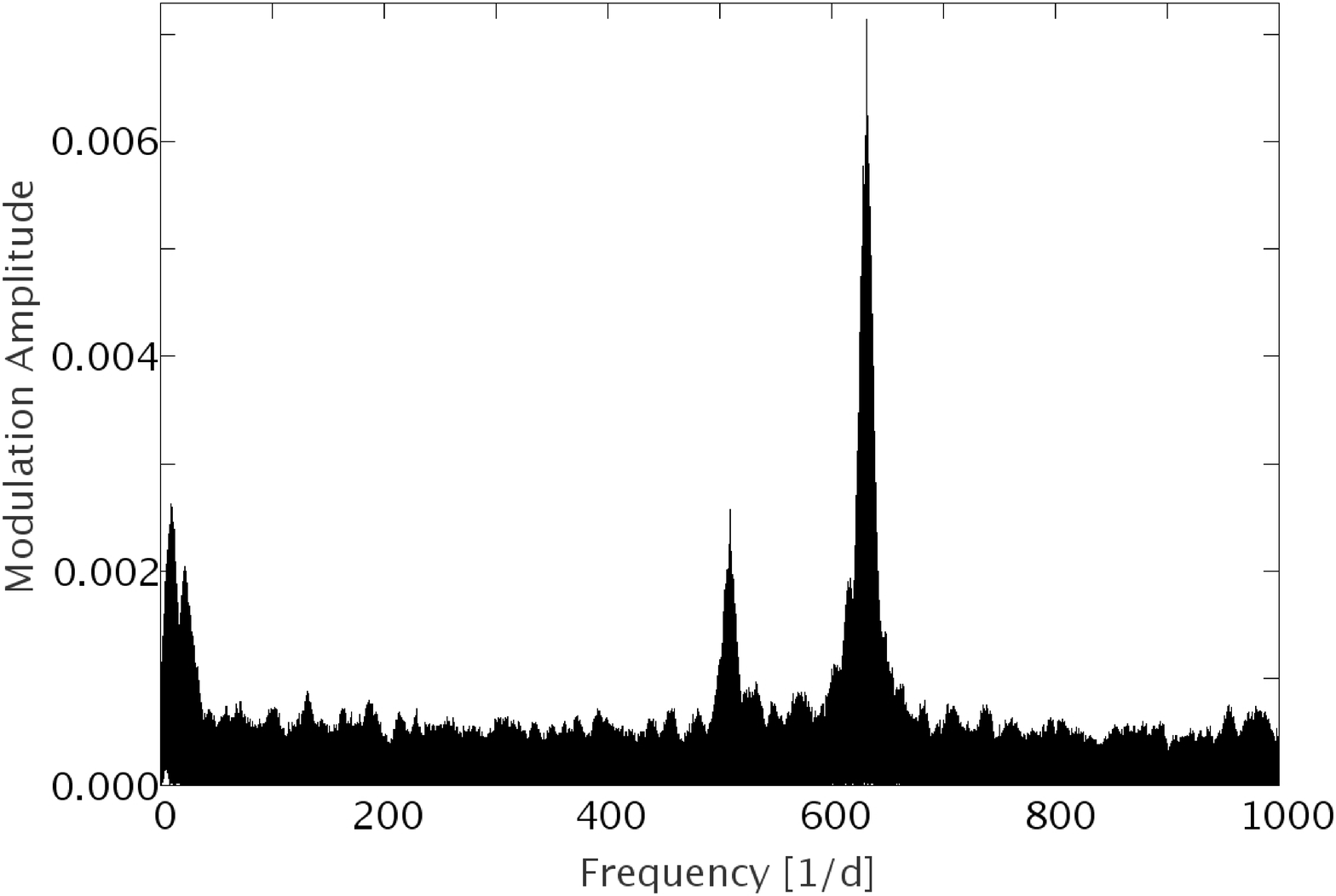}
  \includegraphics[width=.499\textwidth]{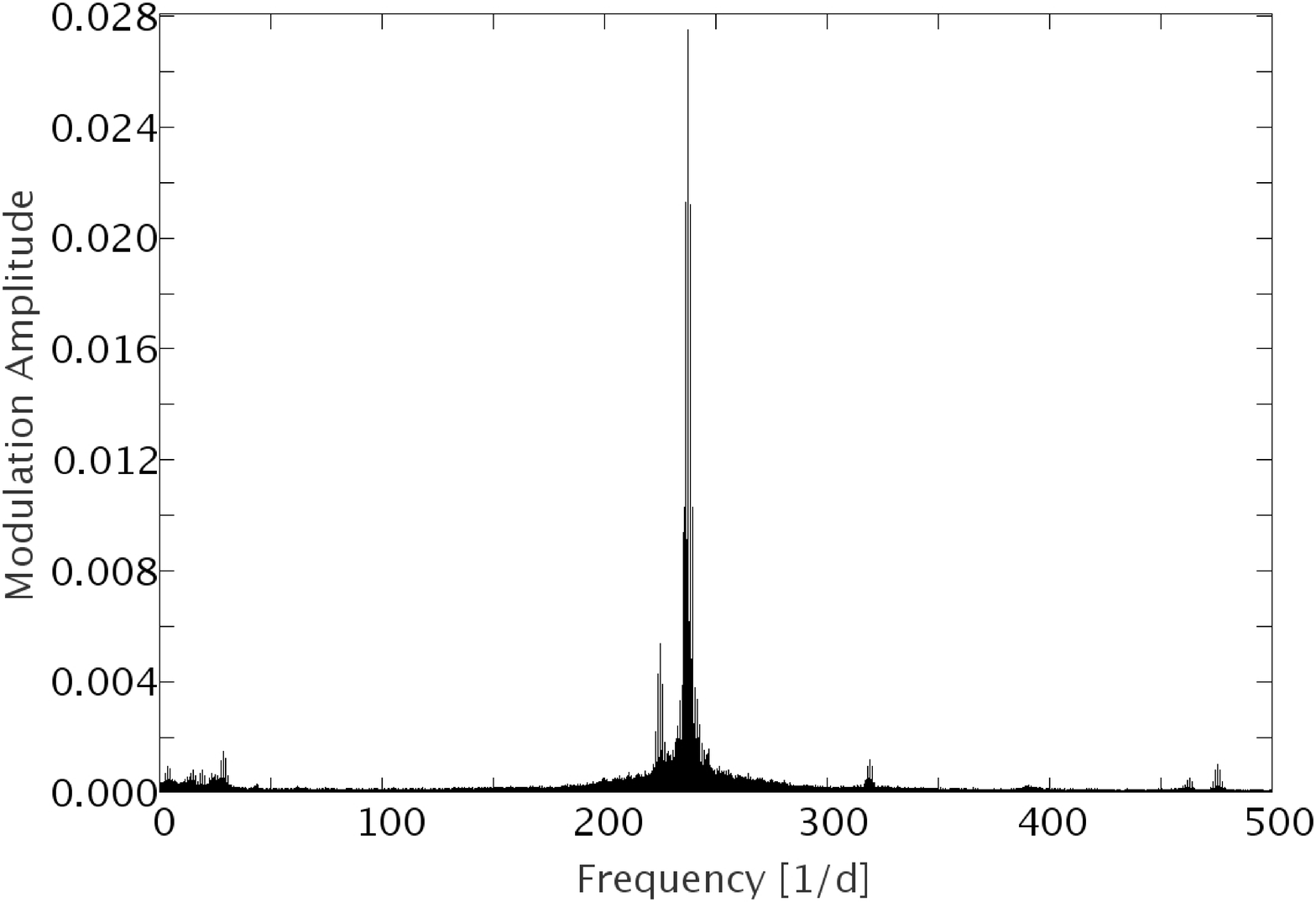}
  \caption{Fourier transforms of our full data sets of HS\,0444+0458
    (left) and HS\,0702+6043 (right). The total  
  time base is 19 months and 28 months, respectively. See more details in Table \ref{tab:a}.}
  \label{fig:34}
\end{ltxfigure}

\section{Observations}
The search for exoplanets around pulsating sdB stars is an ongoing program:
long-term photometric observations are organized in the
\emph{EXOTIME} program (EXOplanet search with the TIming MEthod, lead
by R.\ Silvotti \& S.\ Schuh). A recent review on the status is given
by \citet{2010Ap&SS.tmp..130S}. 
A basic factor for a pulsating sdB star to be examined
for substellar companions with the timing method is the characteristic of its
periodogram. A suitable target shows not too many frequencies, moderately 
large and stable amplitudes and a suitable magnitude. Most importantly,
the pulsations must adhere to phase coherence to a sufficient degree. 
The targets considered here have been tested to fulfill these conditions.
HS\,0702+6043 (Fig. \ref{fig:34}) is a hybrid sdB pulsator (sdBV$_{rs}$) showing a main pulsation 
period of 382s with an amplitude around 27\,mmag \citep{2002A&A...386..249D,2008ASPC..392..339L}.
HS\,0444+0458 (Fig. \ref{fig:34}) is a p-mode pulsator (sdBV$_r$) with a main pulsation period 
of 137s with an amplitude of 7\,mmag \citep{2001A&A...378..466O}.
\par
To derive a single point for the O--C diagrams of our
selected targets, observations during on average at least three to four 
consecutive nights with at least two to three hours per target 
per night are required. The minimum time base of three nights for each block 
is needed to 
resolve the pulsational frequencies. At least three hours per
night are needed for each object to sample enough pulsational cycles. Since
we look at low-amplitude, short-period p-modes (some mmag on a timescale 
of about five minutes or less) we require both a short sampling time of less 
than about 30s \textit{and} very high signal to noise during these short
exposures to detect the low amplitudes. \emph{EXOTIME} conducts regular long-term
time-series photometry using several small- to medium-class telescopes. A
current list of our photometric data archive for the targets HS\,0444+0458 and
HS\,0702+6043 is given in Table \ref{tab:a}.
\begin{table}[t]
\begin{tabular}{lrr}
\hline
    \tablehead{1}{r}{b}{Site and mirror size\rule{2mm}{0mm}}
  & \tablehead{1}{r}{b}{HS\,0444+0458\\Aug\,2008\,--\,Feb\,2010}
  & \tablehead{1}{r}{b}{HS\,0702+6043\\Nov\,2007\,--\,Feb\,2010} \\
\hline
Mt. Bigelow 1.56m &  & 424 h \\
Calar Alto 2.20m & 32 h & 116 h \\
MONET/N 1.20m & 16 h & 98 h \\
Tue 0.80m &  & 62 h \\
Goe 0.50m &  & 53 h \\
LOAO 1.00m & 16 h & 43 h \\
TNG 3.60m & 7 h &  \\
Asiago 1.82m &  & 20 h \\
Loiano 1.52m &  & 14 h \\
Konkoly 1.00m &  & 14 h \\
St. Bok 2.20m &  & 12 h \\
Moletai 1.60m &  & 8 h \\
NOT 2.56m &  & 1 h \\
\hline
total & 71 h & 865 h \\
\hline
\end{tabular}
\caption{Photometric data for HS\,0444+0458 and HS\,0702+6043}
\label{tab:a}
\end{table}

\section{Data Reduction}
After extracting relative light curves in a standard aperture
photometry procedure, detrending and normalization,
it is also absolutely crucial to do the time correction for
Earth's motion in the Solar System properly (the reference point is
the barycenter of the Solar System). In addition, one needs to be
aware of leap seconds. At this stage, there are the relative
normalized light curves given in barycentrically 
corrected time (BJD) versus relative intensity.
\par
Since data from different telescope sites are of different quality, we apply
a weighting scheme to the light curves. Point weights $w$ are assigned to each light
curve data point according to $ w = \frac{1}{\sigma^2}$, with $\sigma$ being the standard
deviation calculated on a subset of $n$ data points, centered on each time stamp, after
having subtracted the pulsation frequencies. This
can be described as sort of a "moving standard deviation". $n$ is an integer, usually
corresponding to the number of data points within two or three pulsational cycles.
Gaps are also taken into account. We use a routine provided by one of us (RS).

\section{Timing method and O--C diagram}
The O--C analysis as a particular application of the timing method is
an approach to measure the phase variations of a periodic function. 
In our case, these periodic functions are the pulsations of the sdB
star leading to periodic intensity variations. Other applications
include the timing of Pulsar signals and the timing of eclipses in eclipsing binary systems.
In all cases the periodic signal is used as a clock and one wants to examine the behaviour
of this clock as a function of time. To do so, it is reasonable to compare the calculated
("C"-part of the O--C diagram) light curve solution of the whole data set with the
actually observed ("O"-part of the O--C diagram) light curve solutions of temporal
subsets of the whole data set.
\par
\begin{figure}
  \includegraphics[height=.45\textheight,angle=90]{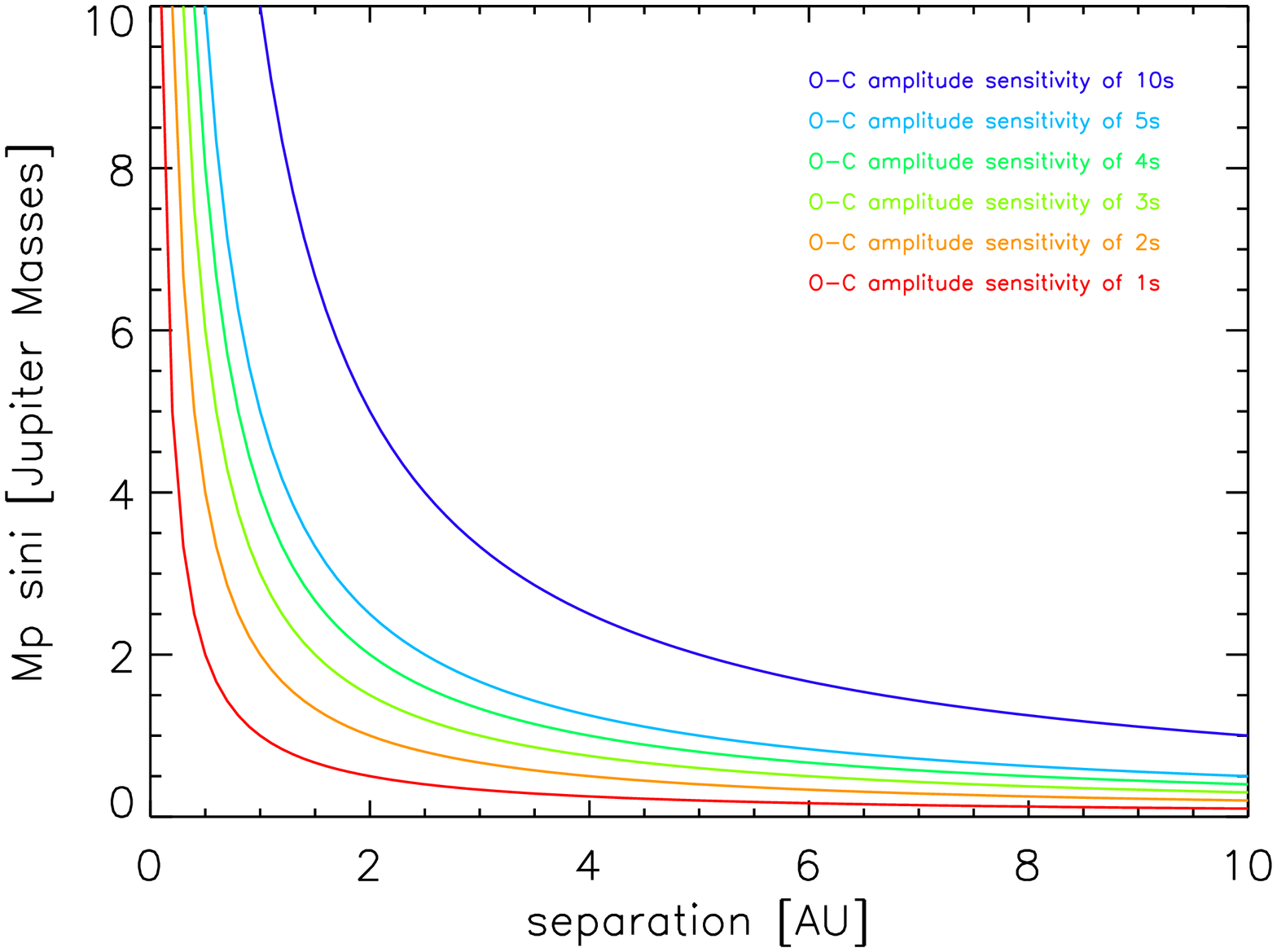}
  \caption{Sensitivity of the timing method applied to pulsating sdB stars. We assume different
  amplitude sensitivities ranging from 10s (uppermost line) to 1s (lowermost line). The 
  justification of the assumed amplitude sensitivities mainly depends on the total time base 
  of observational data being available.}
  \label{fig:2}
\end{figure}
In a first step, a light curve solution is determined for the full light curve.
Since sdB stars are multi-mode pulsators, the solution will consist of a frequency, 
amplitude and phase for each pulsation considered. The frequency solution of this full 
light curve (the "C" part) is then kept fixed and each seasonal light curve is being fitted 
with that fixed frequency, while determining the amplitudes and phases of the seasonal light 
curves (the "O" parts). It is sometimes also be possible to keep both
the frequency and the amplitude of the full 
light curve fixed. For each seasonal light curve the phase difference to
the full light curve can now be determined. These phase differences are translated to
time differences, which, plotted as a function of time, yield an O--C
diagram.
\par
Assuming a reflex motion of the sdB caused by a substellar companion
(translating into a sinusoidal component in the O--C diagram), one can
find a relation between the O--C amplitude $\tau$ and the orbital
separation $a$.  
Given the assumptions of a two body case, a circular orbit, a canonical sdB mass
of 0.5 solar masses, $M_p\ll M_{\star}$, and scaling to Solar System units, one obtains
\begin{equation}
M_p \cdot \sin{i} = \tau \cdot \frac{1}{a},
\end{equation} 
with the companion's mass $M_p [M_{Jup}]$, the inclination $i$ and orbital separation $a [AU]$ . 
The O--C amplitude $\tau [s]$ is
a quantity measured in the O--C diagram, such that assuming different given amplitude sensitivities,
one can plot the minimum detectable planetary mass versus the orbital
separation. This simple estimation has been done 
for different amplitude sensitivities in Fig. \ref{fig:2}. As can be seen, the timing method applied to 
pulsating sdB stars is more sensitive towards larger separations.
\par



\begin{ltxfigure}
  \includegraphics[height=.33\textheight,angle=90]{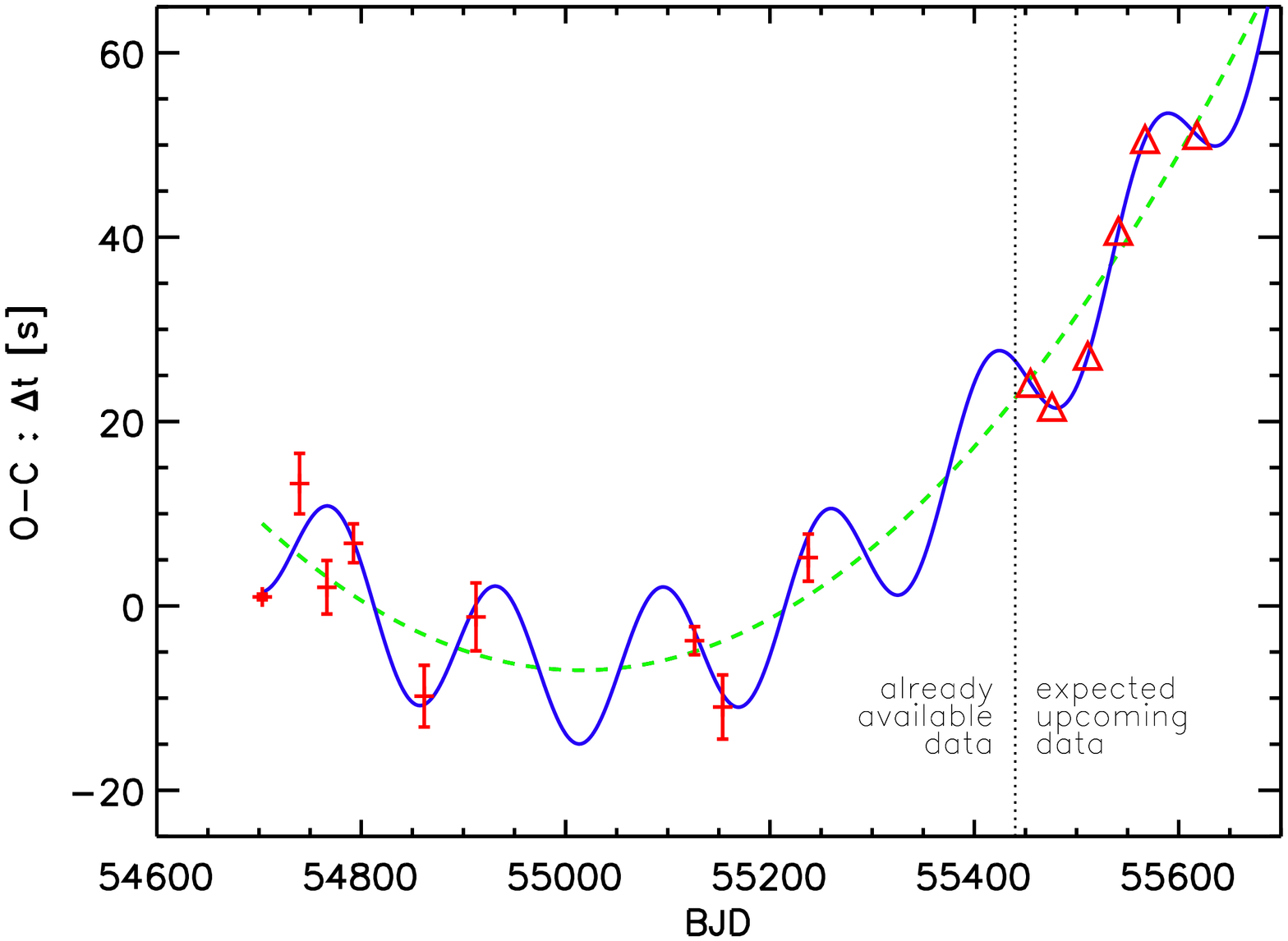}
  \includegraphics[height=.33\textheight,angle=90]{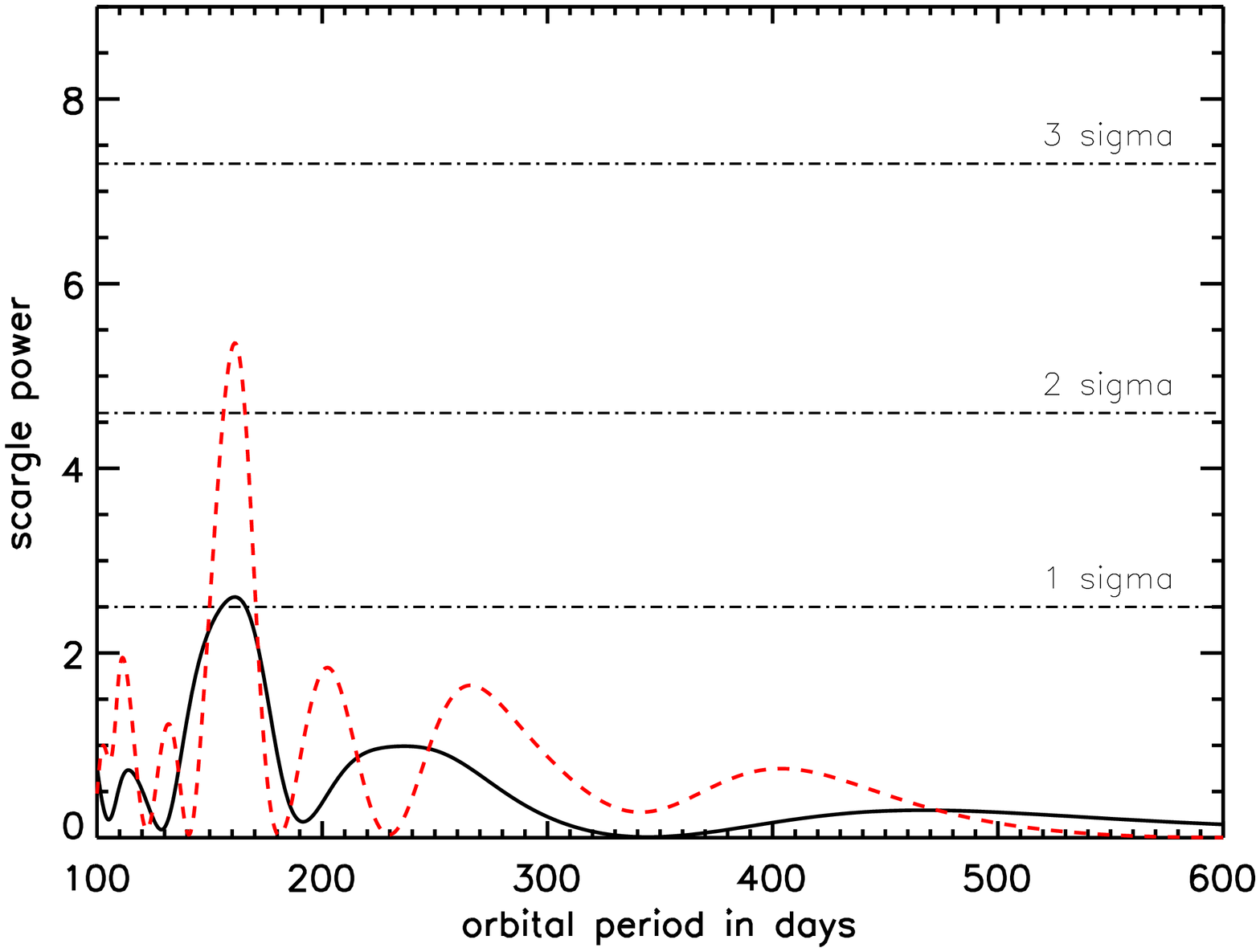}
  \caption{O--C diagram (left) for the main frequency f1 of 
HS\,0444+0458. The symbols with error bars are seasonal data currently available. 
The dashed line is a parabolic component yielding a $\dot{P}$ of
$6.00\cdot10^{-12}$\,d/d or an evolutionary timescale of 0.72 Myr. We
proceed to specifically look for sinusoidal
residuals, which might hint at a reflex motion of a possible companion. The solid line in
the right panel shows a periodogram of the residuals based on the current data
with a peak around 160 days at a significance of only 1$\sigma$. 
Boldly assuming an orbit of 160 days, we calculate the expected signature in the O--C 
(solid line in the left panel). The triangles refer to future observations. 
The predicted effect of the increased time baseline due
to the future observations on the significance in the residual periodogram
is shown as the dashed line in the right panel. If upcoming data support our
current model, the significance in the residual periodogram would rise
above 2$\sigma$, strengthening the hypothesis of a substellar
companion candidate (not currently claimed as real).}
\label{fig:56}
\end{ltxfigure}

\begin{ltxfigure}
  \includegraphics[height=.33\textheight,angle=90]{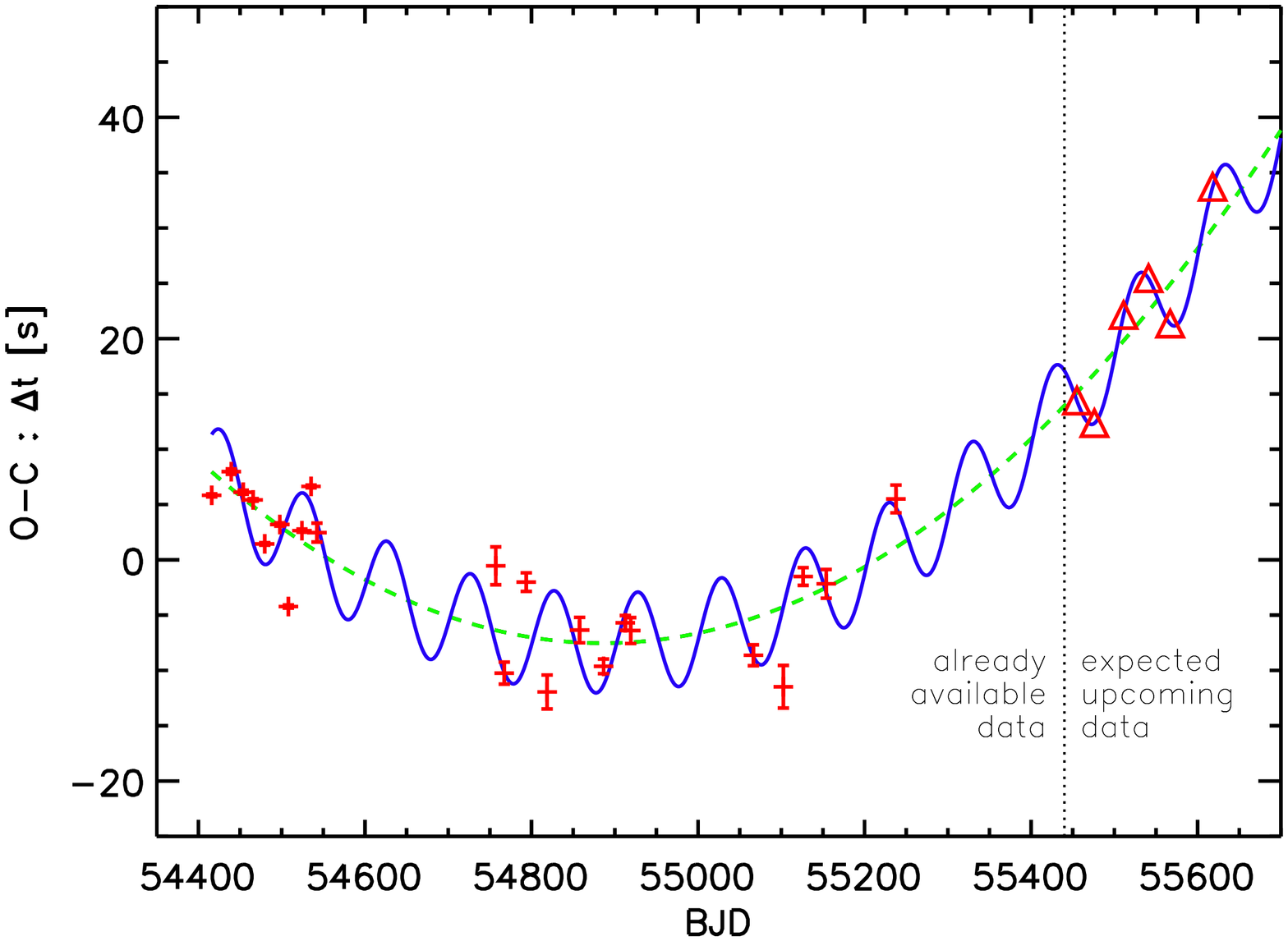}
  \includegraphics[height=.33\textheight,angle=90]{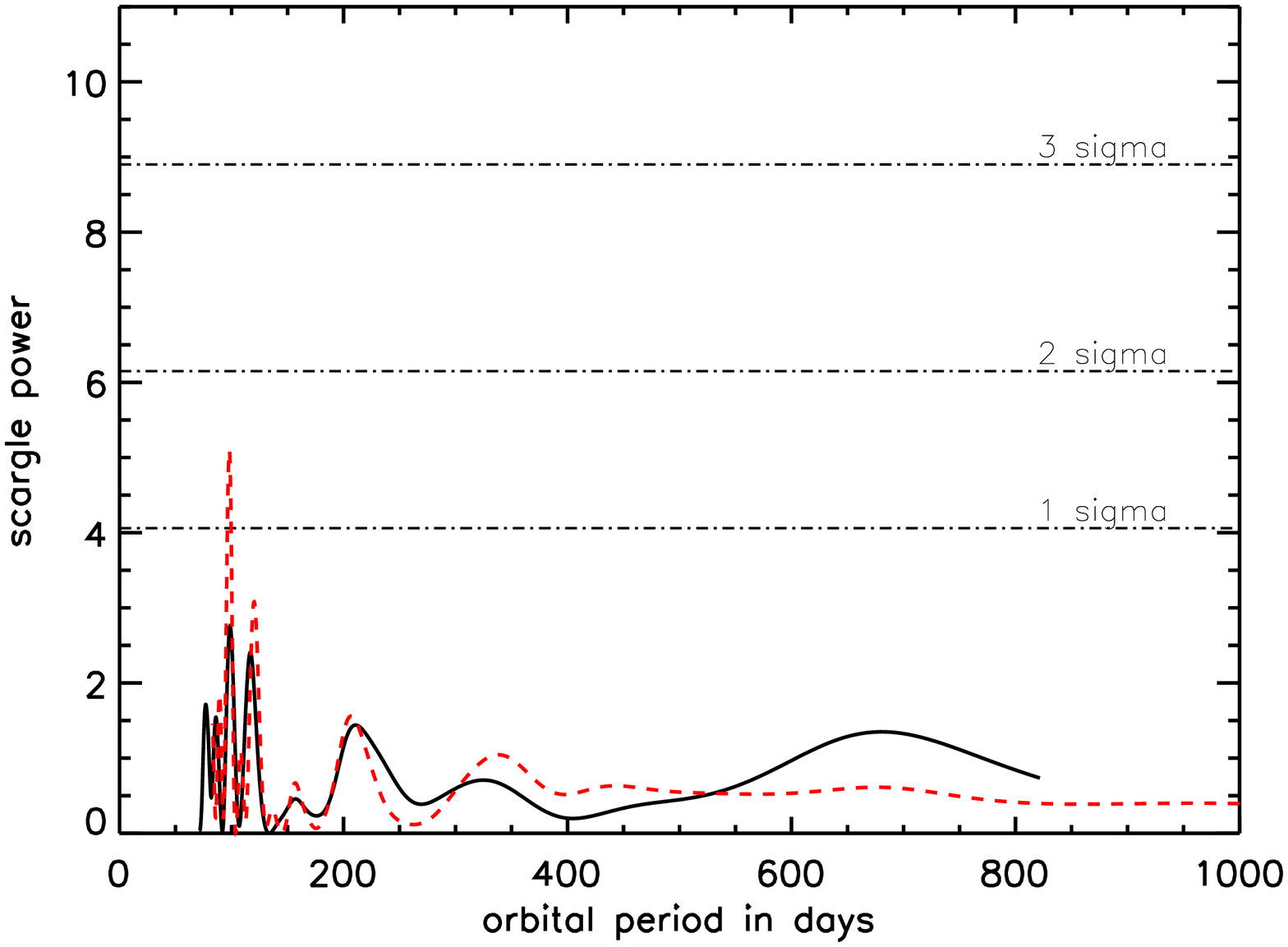}
  \caption{Same as Fig. \ref{fig:56}, but for the main frequency f1 of HS\,0702+6043. 
  The parabolic component yields a $\dot{P}$ of $6.81\cdot10^{-12}$ d/d, translating to an 
  evolutionary timescale of 1.69 Myr. The residuals reveal no significant orbital signal so far.}
\label{fig:78}
\end{ltxfigure}

\begin{ltxfigure}
  \includegraphics[height=.33\textheight,angle=90]{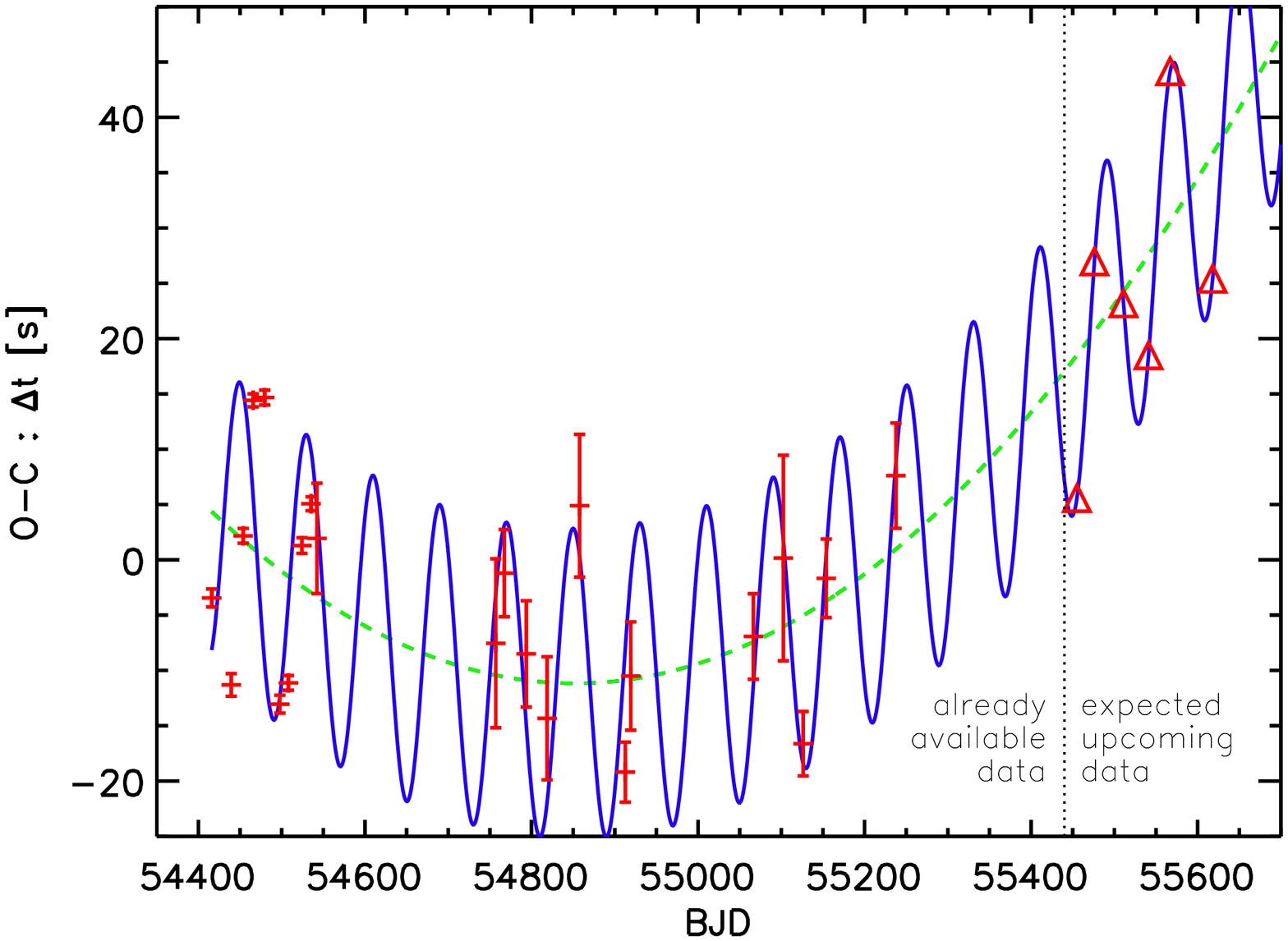}
  \includegraphics[height=.33\textheight,angle=90]{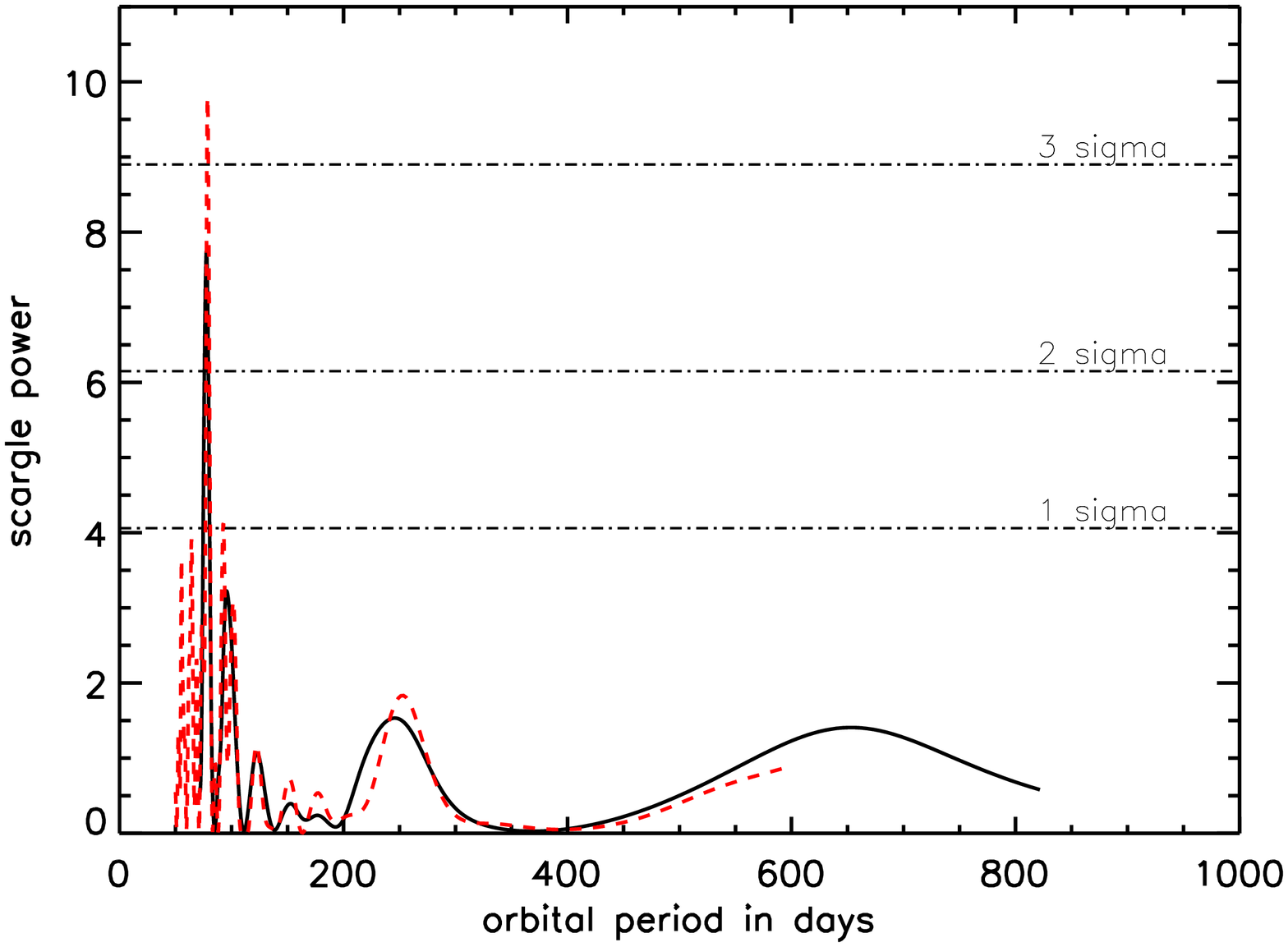}
  \caption{Same as Fig. \ref{fig:78}, but for the second frequency f2 of HS\,0702+6043. 
  The parabolic component yields a $\dot{P}$ of $8.39\cdot10^{-12}$ d/d, translating to an 
  evolutionary timescale of 1.45 Myr. The significant peak close to 3$\sigma$ in the residual
  periodogram mimics a possible companion in a 80 days orbit, but is most probably due to a beating
  of close unresolved frequencies. These two close frequencies have an amplitude ratio close to 2:1
  and can only be resolved in the full data set, but not in the shorter seasonal subsets.}
\label{fig:910}
\end{ltxfigure}

\section{Results}
To get independent results, we conducted the O--C analysis not only 
for the strongest frequencies (f1) in HS\,0444+0458 and HS\,0702+6043, 
but also for the second strongest ones (f2).
Table \ref{tab:b} gives a summary of the frequencies used for the
analysis and our results for $\dot{P}$ and 
the evolutionary timescales. The O--C diagram of the strongest
frequency f1 in HS\,0444+0458 is shown in Fig. \ref{fig:56}. 
Although we find much more pulsations in
the FT of the full data set of HS\,0702+6043, 
these are too weak to be included in the detailed O--C analysis.
O--C diagrams for f1 and f2 of HS\,0702+6043 are visualized in Figs. \ref{fig:78} and
\ref{fig:910}.
\par

\begin{table}[h]
\begin{tabular}{lrrrrr}
\hline
  & \tablehead{1}{r}{b}{Label}
  & \tablehead{1}{r}{b}{Frequency [1/d]}
  & \tablehead{1}{r}{b}{Amplitude [mma]} 
  & \tablehead{1}{r}{b}{$\dot{P}$ [d/d]} 
  & \tablehead{1}{r}{b}{$T_{evol}$ [Myr]} \\
\hline
HS\,0444+0458 & f1 & 631.735086 & 7.13 & $6.00\cdot 10^{-12}$ & 0.72 \\
       & f2 & 509.977714 & 2.53 & $5.27\cdot 10^{-12}$ & 1.02\\
\hline
HS\,0702+6043 & f1 & 237.941086 & 27.51 & $6.81\cdot 10^{-12}$ & 1.69 \\
       & f2 & 225.158793 & 5.60 & $8.39\cdot 10^{-12}$ & 1.45 \\
\hline
\end{tabular}
\caption{Frequencies used for the O-C analysis and evolutionary timescales of the corresponding pulsations}
\label{tab:b}
\end{table}

\section{Discussion}
One goal of the \emph{EXOTIME} program is to determine the evolutionary timescales of pulsating sdB stars.
Here, we considered two particular objects and analyzed the currently available photometry covering baselines
of 19 and 28 months. The $T_{evol}$ values found for both modes f1 and f2 in HS\,0702+6043 and HS\,0444+0458 are 
compatible with current models \citep{2002ApJS..140..469C}. Our results are also compatible with those found 
for HS\,2201+2610 (V\,391 Peg): \citet{2007Natur.449..189S} find a $\dot{P}$ of $1.46\cdot 10^{-12}$ and
$2.05\cdot 10^{-12}$ for the two strongest frequencies, corresponding to evolutionary timescales of
7.6 Myr and 5.5 Myr, respectively.
\par
The second part of the \emph{EXOTIME} program, examining the targets for substellar companions, does not
come to a conclusion yet. The significance of any periodic signals in
the residual periodograms is too low, but we make testable predictions for HS\,0444+0458.
The particular strength of the timing method, its ability to uncover planets in
wide orbits, at the same time means that long timescales are involved:
for detecting a planet in a \emph{x}-year orbit one obviously needs
a time base of more than \emph{x} years to uncover it.
The timing method is therefore very consuming in terms of 
observing time if applied in a way to really play its strength. 
Provided that we will regularly gather more data, we 
expect to give a detailed statement on the possibility of the presence of substellar companions within 
the next year.


\begin{theacknowledgments}
RL, SS and RS want to thank all observers who contribute to the \emph{EXOTIME} program.
RL appreciates financial support by the organizers who waived the conference fee.
\end{theacknowledgments}






\hyphenation{Post-Script Sprin-ger}

\end{document}